\documentclass[amsmath,amssymb]{revtex4}

\usepackage{graphicx}
\usepackage{dcolumn}
\usepackage{bm}

\usepackage{amsfonts}
\usepackage{graphicx, color}
\usepackage{hyperref}

\begin{document}


\title{Reply to Andrew Hodges}


\author{Tien D. Kieu}
\email{kieu@swin.edu.au}
\affiliation{Centre for Atom Optics and Ultrafast Spectroscopy, ARC
Centre of Excellence for Quantum-Atom Optics, Swinburne University
of Technology, Hawthorn 3122, Australia}



\date{March 1, 2006}

\begin{abstract}
We separate the criticisms of Hodges~\cite{Hodges2005} and others
into those against the algorithm itself and those against its
physical implementation. We then point out that {\em all} those
against the algorithm are either misleading or misunderstanding, and
that the algorithm is self consistent. The only central argument
against physical implementations of the algorithm, on the other
hand, is based on an assumption that its Hamiltonians cannot be
effectively constructed due to a lack of infinite precision.
However, so far there is no known physical principle dictating why
that cannot be done.  To show that the criticism may not be a
forgone conclusion, we point out the virtually unknown fact that, on
the contrary, simple instances of Diophantine equations with
apparently {\em infinitely precisely} integer coefficients have {\em
already} been realised in experiments for certain quantum phase
transitions. We also speculate on how central limit theorem of
statistics might be of some help in the effective implementation of
the required Hamiltonians.
\end{abstract}

\maketitle

\section*{Introduction}
Since our first proposal in 2001 of an algorithm employing quantum
adiabatic processes to render the classically noncomputable
Hilbert's tenth problem into that of a quantum mechanically
computable~\cite{kieu-intjtheo,kieu-contphys,kieu-royal,kieuFull},
there have been a lot of interests as well as criticisms. That
situation is inevitable for such seemingly radical a claim against
the accepted and accustomed wisdom.	 On the other hand, it is also a
natural and healthy state of scientific research wherein all new
proposals must be subjected to rigorous examination against what is
currently known.	In our case, however, most of the criticisms are
not in the formal literature but only in informal discussions in
private or on internet forum and at the many seminars and conference
presentations that we have given. We have tried to collect all the
criticisms known to us and evaluate them one by one, for instance,
in one of our recent postings~\cite{Kieu2005}.

This paper is to analyze all the points raised by Andrew
Hodges~\cite{Hodges2005}. We first note that most of those points
have already been raised before by others to which we have already
discussed and addressed accordingly, for example in~\cite{Kieu2005}.
But as Hodges has also included few new objections among them, we
will address all of them here again in some details.

We will make the distinction between the algorithm itself and its
physical implementation in the below.

Against the algorithm, all the points, except one, raised by Hodges
are just misunderstandings, which we will try to clarify in the next
Section. The only exception is Hodges's reference to a paper by
Smith, who has provided a counterexample~\cite{Smith2005} to our
previous claim about a sufficient condition for a criterion that
employs measurement probability to identify the final true ground
state among the infinitely many available states in the quantum
adiabatic algorithm. We have investigated the counterexample and
identified the cause for the insufficiency of our previously stated
condition~\cite{kieuNew}. Subsequently, we have given a mathematical
proof~\cite{Kieu2006} for the same criterion, validating the
mathematical and logical consistency and correctness of our proposed
algorithm. We will not repeat those arguments here.

With respect to the question of physical implementation of the
algorithm, Hodges has reiterated once again the issue of infinite
precision which has been raised elsewhere by others before. We will
discuss our responses~\cite{Kieu2005} to that issue further in more
details in the second last Section.

\section*{Fundamental concepts for a basis for further discussions}
We should note firstly that there is a whole hierarchy of the
noncomputables~\cite{Rogers1967}; that is, some are more
`noncomputable' than others.	Computability of certain
noncomputable, if could be ascertained, does not mean that {\em all}
the noncomputables are then computable. Our claim of (quantum)
computability is restricted in that sense, it is only applied to
Hilbert's tenth problem, or equivalently the Turing halting problem,
or any equivalent problem. (Note also that such equivalence is not
trivial at all; it took more than 70 years and generations of
mathematicians to establish the equivalence of Hilbert's tenth
problem and the Turing halting problem.)

What constitutes, in particular, the noncomputability of Hilbert's
tenth problem? For each Diophantine equation without any parameter,
there is nothing noncomputable about whether that particular
equation has any integer solution or not.	 There always exists a
finite mathematical procedure to find the answer for such a question
for any single Diophantine equation (but not always with a {\em
family} of parametrised Diophantine equations). If we have not yet
had a (recursive) procedure to obtain that answer then it does not
mean that such a procedure is ever out of reach.  In that regards,
noncomputability is similar to the concept of randomness; there is
nothing random about a {\em single} bit or number by itself.  (The
notion of randomness only applies to a series of such bits or
numbers which does not have any statistical pattern or, if we appeal
to algorithmic information theory, does not have a pattern which can
be encoded more effectively and economically than the length of the
series.)

Nor there is anything noncomputable about a collection of {\em
finitely} many Diophantine equations, because we can always
concatenate all the procedures for all the equations (as there is
always a procedure in principle to determine the existence of
solution for each equation) into a {\em finitely} collective
procedure that can be applied to the whole collection.

What constitutes the noncomputability of Hilbert's tenth problem is
the fact that we ask for a {\em single} finite procedure which can
be applied to all the elements of sets of {\em countably} many
Diophantine equations. The application of Cantor's diagonal
arguments to the Turing hating problem, see for
example~\cite{OrdKieu-diag}, has established that were there a
finite and recursive Turing machine that can be applied to determine
the halting of, or lack of, any Turing machine then contradiction
and inconsistency would have to follow. What that means for
Hilbert's tenth problem is that there is {\em no} single finite
recursive algorithm for {\em all} Diophantine equations, but for
each given equation we have to find a recursive algorithm {\em anew}
each time.

We could confirm the fact, which Hodges has also mentioned, that if
we could resolve Hilbert's tenth problem in the positive then we
would be able to resolve not only the Riemann hypothesis but also
the Goldbach conjecture, Fermat's last theorem, and the four-colour
problem, to name a few. And we would be able to do that with a {\em
singly unified} approach too.	 This is because these problems belong
to a class of problems each of which has a Diophantine
representation -- namely, the resolution of each problem can be
rephrased in terms of the existence or lack of solution of some
Diophantine equation corresponding particularly to that specific
problem.  (On the other hand, as there is nothing sacred or
noncomputable about a single parameter-free Diophantine equation,
the resolution of Hilbert's tenth problem has no other consequence
otherwise on the individual equation.)

To illustrate that Cantor's arguments cannot rule out
hypercomputation (for a general discussion on this issue,
see~\cite{OrdKieu-diag}), we have pointed out that the probabilistic
arguments of our algorithm can avoid the usual Cantor's arguments
and thus leads to no contradiction.	 Nor there would be any
contradiction if our quantum algorithm cannot be encoded with a
G\"odel integer. And this is how we could reconcile the quantum
adiabatic algorithm with Cantor's diagonal arguments in a consistent
manner. Certainly the probabilistic nature of the algorithm is not
sufficient to imply that it is hypercomputational. For that, we have
had to construct and show explicitly that it could, as a single
algorithm, resolve any instance of Hilbert's tenth problem.	 A
criticism, as one of Hodges', on the probabilistic nature of our
algorithm not only misses the point but also is not supportable.

We always have explicitly stated and repeatedly emphasized the
probabilistic nature of our algorithm in that it can determined up
to any predetermined probability whether or not a particular
Diophantine equation has any solution. That probability can be
chosen arbitrarily close to one, but can never be one (except in the
event the equation has a solution, in which case we could verify by
substitution). If Hodges and other mathematicians are not willing to
accept that as a way to nonrecursively resolve Hilbert's tenth
problem then we would not be able to share any common ground at all,
and perhaps no further exchange is possible. If that is the case, we
still must point out, as a parting remark, that there is as yet {\em
no} classical recursive algorithm which can resolve the problem in
the same probabilistic manner as the quantum algorithm.

\section*{The questions of various infinities in the algorithm}
Hodges and others often are puzzled by the apparent ability of the
quantum algorithm to {\em explore an infinite space in a finite
time}. That puzzlement comes from a simplistic expectation that in
order to determine whether a Diophantine equation has any solution
or not one would have to explore the {\em whole} integer space. Such
an exploring task apparently could not be accomplished in a finite
time in the case the equation has no solution at all. As a matter of
fact, the expectation is too simplistic and not quite correct. Were
it correct then we would easily be able to, and certainly not need
the sophisticated Davis-Putnam-Robinson-Matiyasevich
theorem~\cite{hilbert10} and then Cantor's diagonal arguments, to
establish the noncomputability of Hilbert's tenth problem.	The fact
of the matter is for each Diophantine equation we only need to
explore a {\em finite} domain in the space of appropriate tuples of
integers; the equation has a solution {\em if and only if} the
solution resides in that finite domain (bounded by the so-called
test function, see Davis~\cite{Davis1976}). This property is quite
remarkable and is applicable to a wider class of so-called {\em
finitely refutable} problems, see a theorem
in~\cite{CaludeBookRandom}.  Once such a finite domain is known for
a Diophantine equation, it is just a matter of substitution of a
finite number of integers to determine if the equation has any
solution at all. The noncomputability of Hilbert's tenth problem
comes from the fact that there can be no {\em single} finite
recursive procedure to determine such a finite domain for each and
every Diophantine equation.  There must be, in other words, at least
one Diophantine equation that is not susceptible to the treatment of
any given recursive procedure.

On the other hand, we claim that a single quantum adiabatic
procedure, as described by the quantum algorithm, can be applied to
each and every equation.  Surely, in a finite time the quantum
mechanical (normalisable) wave function can only spread out from its
initial wave form to explore an effectively finite domain of the
underlying Hilbert space.  But the domain so explored is the
relevant domain sufficient for the purpose of finitely refuting or
confirming the existence of solution for the equation.  The finitely
refutable character of Hilbert's tenth problem manifests itself in
the quantum algorithm as the finiteness in the energy of the final
ground state and in the time that this ground state can be obtained
and identified (by our identification criterion via a probability
measure~\cite{Kieu2006}).

To drive home the point that only a finite portion of a space can be
explored in a finite time, a fact with which we do agree, Hodges has
also mentioned Grover's search algorithm~\cite{Grover} in an
unstructured database to erroneously imply that quantum adiabatic
computation cannot search more efficiently than the time complexity
offered by Grover's algorithm (which is of the order of the square
root of the database size) and thus cannot compute Hilbert's tenth
problem -- a statement with which we disagree. Such a statement is
misleading and incorrect in at least three aspects. Firstly, quantum
adiabatic computation could accomplish the search in a time {\em
independent} of the size of the database, provided sufficient energy
must be supplied, see the
references~\cite{Eryigit2003,Das2003,Wei,Kieu2006b}. The energy,
secondly, need not be proportional to the square root of the
database size, quantum entanglement can reduce substantially the
required energy, as demonstrated in a quantum adiabatic algorithm
for the NP-complete travelling salesman problem~\cite{Kieu2006b}.
And thirdly, for any instance of Hilbert's tenth problem, as we only
need to search in a finite domain appropriate for the specific
parameter-free equation under consideration thanks to the finitely
refutable property, both the energy required and the time taken are
{\em finite} for a successful execution of the quantum algorithm in
the infinite underlying Hilbert space.  Some finite information
about the final ground state is that all we could have, there is no
infinite amount of information here as misunderstood by Hodges.

Metaphorically speaking, the initial ground state of the quantum
algorithm provides one end of a {\em finite} string along which we
can trace to the final ground state at the other end. This can be
done quantum mechanically because of the quantum adiabatic theorem,
which asserts that a particular eigenstate of a final-time
Hamiltonian, even in dimensionally {\em infinite} spaces, could be
found mathematically and/or physically {\em in a finite time}. This
is a remarkable property, which is enabled by quantum interference
and quantum tunneling with complex-valued probability amplitudes,
and in principle allows us to find a needle in a particular infinite
haystack! Such a property is clearly {\em not} available for
classical recursive search in an unstructured infinite space.  Being
built on these principles, our algorithm is far from being a simple
brute-force search as claimed by Hodges.

The applicability of quantum adiabatic theorem here is warranted as
we have shown that~\cite{kieuFull}, for the quantum adiabatic
processes of the algorithm, there is no level crossing in the
spectral flow not only between the instantaneous ground state and
the instantaneous first excited state but also between {\em any}
pair of instantaneous eigenstates.	Interestingly, there are other
versions of quantum adiabatic theorem~\cite{Avron1999} which do not
require non-zero gaps between the energy levels.  We will not need
it here but may need it later to resolve a technical degeneracy
problem for our algorithm, see below.

Incidentally, we have also been able to derive a lower bound on the
computation time $T$ for a general quantum adiabatic computation,
\[4<g(\theta) T \Delta_I E.\]
This lower bound on the computation time incorporates the initial
ground state and the spectrum of the final Hamiltonian together in
$\Delta_I E$, which is defined as the energy spread of the initial
state in terms of the final energy. The manner of the time
extrapolation is further reflected in $g(\theta)$.  This condition
is applicable to finite and infinite spaces, and states that the
more the spread of the initial state in energy with respect to the
final Hamiltonian, the less the lower bound on the running time. It
thus also emphasizes the fact that energy must be considered in the
running of quantum adiabatic computation, and perhaps in all
computation if they all are indeed physical in the end.  Energy,
thus, should be another dimension of algorithmic complexity in
addition to those of time and space.

Because of the finitely refutable character of Diophantine
equations, we do not, contrary to naive expectation, really need an
infinite space to carry out the algorithm for any given
parameter-free Diophantine equation. A finite but sufficiently large
space will do. Once the final ground state has already been included
in this finite space the addition of further Fock states which are
highly excited states of the final Hamiltonian will only change the
dynamics negligibly, as can be inferred from a set of infinitely
coupled differential equations that we have derived
in~\cite{kieu-royal}.

We might or might not be able to devise a procedure according to
which we could tell whether, for a given Diophantine equation, the
underlying Hilbert space even though finite is sufficiently large.
But in order to keep the algorithm simple at this level, we prefer
to employ infinite spaces in the algorithm, which are quite
appropriate for the set of all Diophantine equations (but not quite
necessary because an unbounded underlying space might do).	We have
to point out here again that dimensionally infinite Hilbert spaces
are not exceptions in quantum mechanics, they are, rather, the norm.
Without infinite Hilbert spaces, for instance, we would never have
had any realisation of the fundamental commutator
\[[\,x,p\,] = i\hbar.\]
(To wit, were $x$ and $p$ finite matrices, the mathematical trace of
the lhs would vanish, contradicting the non-vanishing trace of the
rhs!)

The last point of this Section is about possible degeneracy of the
final ground state which originates from possible multiplicity of
the globally minimum value of the square of the Diophantine
equation. We have proposed the use of additional symmetry breaking
terms in the final Hamiltonian to avoid the degeneracy, before
asymptotically removing these terms in some perturbation-theory
treatment in order to recover the original Hamiltonian.  We do not
quite understand Hodges' objection to the application of
perturbation theory and therefore cannot comment on his objection,
except only wish to mention here that we can show that any
degeneracy (even with infinite multiplicity) can indeed be lifted
with the introduction of the symmetry breaking terms.	 A
mathematical proof for this is exactly the same as that for the no
crossing of any pair of instantaneous eigenstates in the spectral
flow we have already mentioned above.

Apart from the proposed use of symmetry breaking terms, the problem
of ground-state degeneracy can certainly be removed if there is
indeed a {\em single-fold} Diophantine representation for every
listable set~\cite{Davis1976}.  The existence of such a
representation is, however, still an open problem at present time.
The same degeneracy problem for our algorithm could perhaps be
removed with a suitable application of the versions of quantum
adiabatic theorem which require no gap between the
levels~\cite{Avron1999}. But this is only a speculation at present
and would require further investigation elsewhere.

\section*{Is the algorithm physically implementable?}
As distinct from the criticisms above on the quantum algorithm, we
now move to the question of its physical implementability.

The issue of infinite precision that Hodges mentioned has in fact
been raised and discussed before~\cite{Kieu2005}. Martin Davis was
the first person who brought to our attention around 2002 that a
lack of precision in an implementation of the integer coefficients
of Diophantine equations, for example, those in the equation
\[x^2 - 2 y^2 =0,\]
would lead to a wrong representation of, and thus to a wrong answer
to, the original equation.  And it is thought that this precision
problem cannot be overcome in any possible implementation of the
quantum algorithm.

Before analyzing the issue, we have to point out here that Hodges'
use of the example of quantum simple harmonic oscillator to dismiss
the physical feasibility of our algorithm is inappropriate.	 The
lack of infinite precision in the characteristic length, or
equivalently in the frequency, of the oscillator is irrelevant here.
We only need to distinguish one quantum from two quanta and in order
to do so we do not need infinite precision in the oscillator
frequency, $\Delta\omega=0$, because it suffices to have $(\Delta
\omega/\omega)\ll 1$.  That is how we are able to count photons, and
confirm the fundamental quantisation nature of light, even though we
do not have absolute precision in their frequencies. Likewise for
our algorithm, sufficiently precise finite precision also suffices
for the purpose of separating the integer-valued energy eigenvalues
of the final Hamiltonian.

Coming back to the issue of precision of the coefficients in a
Diophantine equation, we need to distinguish the infinite precision
from the unbounded precision.	 Infinite precision, the kind of
perfect precision for any digit and all digits of an infinite number
of digits of a measured value, is what Hodges and some other people
claim to be necessary for an implementation of the algorithm.  With
an assumption that such an infinite precision {\em per se} is not
available to our physical instruments, even in principle, they have
come to a negative conclusion accordingly. What irrefutably
available to our instruments, instead and at least in principles, is
the second kind of precision wherein we can obtain an accuracy for
an unbounded but finite number of digits, as many digits as we want
provided we have the time and the resources to do so. There exists
no physical principle against this kind of precision -- even though
we may have to pay the price in, apart from the time and resources
required, some correspondingly worse accuracy for a conjugate
quantity as demanded by some quantum mechanical uncertainty
principle, but that is alright for our purpose.	 Our opinion is that
it may {\em not} be a forgone conclusion, as some may have thought,
that the physically available precision is not adequate for an
implementation of the proposed algorithm.

Firstly, our ultimate objective is to obtain the ground state of a
desired Hamiltonian, and the algorithm is `just' a means to that
end, but a universal means nevertheless.	In saying that the
negative resolution of Hilbert's tenth problem of the (platonic)
mathematical world also carries over to the physical world in such a
way that we could never achieve the above objective is equivalently
to saying that we would never be able either to identify some ground
states, or to construct some suitable Hamiltonians, or both.  That
would have been a very stringent constraint on the physical world
and would have resulted in an entirely new physical principle -- but
we do not have any reason or anything to support this negative
conclusion.

We, as yet, have absolutely no `no-go' physical principles which
dictate that there must exist a physical system which we cannot cool
down to the ground state, or in which there must exist a physical
limit of a distinctively non-zero temperature beyond which we cannot
proceed any further. The third law of thermodynamics, and its
generalisation to the quantum domain, only states that we should not
be able to obtain exact absolute zero temperature, or equivalently,
that we should not be able to obtain a ground state with total
certainty, but that is fine as we do not need to obtain a ground
state with unity probability in order to identify it -- see our
probability criterion for ground state
identification~\cite{Kieu2006}.  But the third law does not demand
that there must be a non-zero lower bound on achievable temperature,
or equally, that there must be an upper bound of obtainable
probability for a ground state. (If anything, the postulate of
projective measurement in standard quantum mechanics may contradict
some classical statement of the third law; we may discuss this
interesting issue elsewhere.)

The experimentally confirmed phenomenon of the {\em thermal} phase
transition of Bose-Einstein Condensation in some kind of traps, for
example, otherwise demonstrates that collective quantum effects do
enable us to obtain and identify some particular ground state with
arbitrarily high probability and achieve temperature arbitrarily
close to absolute zero.

Likewise, there exists no known physical principle against the
possibility of effective construction of any desirable Hamiltonians.
In fact, an effective realisation of the quantum adiabatic algorithm
for the special case of linear Diophantine equations has already
been achieved recently~\cite{Greiner02}, even though it is only
known by others as the {\em quantum} phase transition of superfluid
to Mott-insulator. We refer the readers to the Appendix for some
detailed discussion of our claim that this interesting and seemingly
unrelated quantum phenomenon is indeed connected to the quantum
algorithm.	In a few words, quantum phase transition is different
from thermal phase transition in that, while the latter originates
from the {\em thermal} fluctuations due to the competition between
the opposing requirements of minimising the energy on the one hand
and of maximising the entropy on the other, the former phase
transition originates from the {\em quantum} fluctuations due to
another kind of competition between opposing quantum mechanically
conjugate terms in the same Hamiltonian.	The particular superfluid
- Mott insulator transitions are captured by the Bose-Hubbard model
which contains as apart of the Hamiltonian the term
\[ (a^\dagger a)^2 -2\tilde m (a^\dagger a) + \tilde m^2,\]
and thus can be viewed as some physical realisations of instances of
Hilbert's tenth problem, namely the Diophantine equations,
\[x - \tilde m =0.\]
This fact is surprising, extraordinary and of great consequence; and
will be explored further elsewhere.

Granted that this is a very simple type of Diophantine equations,
but our point here is to emphasize that we should not rule out the
possibility of effective implementation of the quantum algorithm,
because we may be able to count on the help of collective quantum
mechanical behaviours as in the quantum phase transitions above or
of something else.

Is there any other way that unbounded precision may suffice in
general? We speculate that the Central Limit Theorem of statistics
might be of some use here. Let us illustrate the idea with the first
Diophantine equation above, and let us denote its lhs by
$P(u,v;x,y)$,
\[P(u,v;x,y) = ux^2 - vy^2.\]
In realistic implementation, we can treat the parameters $u$ and $v$
as random variables with some {\em finite} variances and aim to have
the averages $\bar u = 1$ and $\bar v = 2$.  In an instance of
implementation, the parameters take some random values $(u_i,v_i)$,
which may or more likely may not be $(1,2)$, resulting in
$P(u_i,v_i; x,y)$ as the input for the quantum algorithm, and
consequently in a subsequent randomly distributed output
$(x_i,y_i)$, which is the location of the corresponding ground
state. Central limit theorem, if applicable, would then ensure that
the average, for a large number of repetitions $N$, of such outputs
$(x_i,y_i)$ from our algorithm should tend to a central value
corresponding to the output from the input $(u,v) = (\bar u, \bar v)
= (1,2)$. As the average of these outputs should have a statistical
spread that is inversely proportional to $\sqrt N$, we may be able
to reduce this variance, by having sufficiently large $N$, until we
could confidently confirm the (integer-valued) central value, which
should be effectively the output for $P(1,2;x,y)$.

While it is certainly less restrictive to assume infinite precision
for the average values of the coefficients rather than the
coefficients themselves, it is still unclear whether this assumption
on the average values could be satisfied. (The in-principle
unsatisfiability of this, on the other hand, would imply that we can
never, as a matter of fundamentals, get rid of systematic, as
opposed to statistical, errors.)  Note also that in order to apply
the central limit theorem we need to have a {\em finite} variance
for the algorithm outputs.  But, as the outputs $(x_i,y_i)$ for
randomly distributed inputs $(u_i,v_i)$ may not be bounded from
above, it is yet to be shown that the former could have a finite
variance even if the latter does.

Obviously, this speculation of ours requires further investigations.

\newpage
\section*{Concluding remarks}
We separate the criticisms of Hodges and others into those against
the algorithm itself and those against its physical implementation.
We then point out that {\em all} those against the algorithm are
either misleading or misunderstanding, and that the algorithm is
self consistent -- which is the best one could do for such a
mathematical entity as the algorithm.

We suspect that all these confusions might perhaps be originated
from a mix-up in thinking that Hilbert's tenth problem is the same
as a search problem in an unstructured and infinite database.  While
the latter unstructured search problem is quite general and
difficult, each parameter-free Diophantine instance of Hilbert's
tenth problem {\em does} provide some structure, which is the
information mathematically encoded in the equation itself, for a
search for the global minimum of its square. Furthermore, the search
space for such a minimum is always finite. To further highlight
possible pitfall for such a simplistic comparison with unstructured
search, let us consider the counterpart of Hilbert's tenth problem
over the real numbers, that is, the question of a single universal
procedure to determine the existence or lack of {\em real} solution
for any given multivariate polynomial with {\em real} coefficients.
Were the brute-force approach of unstructured search in the real
numbers the only one available, one would have concluded that this
would be another noncomputable problem because such a search in an
infinite space of the reals, which is even `larger' in cardinality
than that of the integers, could have never been completed in a
finite time. On the contrary, and perhaps fortunately, such a
conclusion is wrong: Tarski showed that~\cite{Tarski1951}, unlike
Hilbert's tenth problem, this counterpart problem is classical {\em
computable} -- thanks to the structures mathematically encoded in
the polynomials themselves.

Thus, general considerations, as those of Hodges, borrowed from a
general unstructured search not only are inapplicable to our
algorithm but could also be quite misleading.

On the other hand, the only central argument against physical
implementations of the algorithm is based on an assumption that its
Hamiltonians cannot be effectively constructed due to a lack of
infinite precision. To show that this may not be a forgone
conclusion, we draw attention to the virtually unknown fact that, on
the contrary, simple instances of Diophantine equations with
apparently {\em infinitely precisely} integer coefficients have {\em
already} been realised in certain experiments known as quantum phase
transitions! We also speculate on how central limit theorem might be
of some help in the effective implementation of the required
Hamiltonians.

Computability should constitute of both consistency and also
implementability.	 However, the issue of implementability can only
be settled either in the negative by citing prohibiting physical
principles, of which there is none known at present, or in the
positive by actual and general demonstrations, of which there are so
far only special cases connected to certain quantum phase
transitions.  The awaited final outcome must and can only be bounded
by physical laws.	 Until it is settled one way or another, it should
be remembered that premature and prejudiced judgement has never
served us well.

\appendix
\section{Linear Diophantine equations and the superfluid to
Mott-insulator quantum phase transition} The recent experimental
demonstration~\cite{Greiner02} of the superfluid - Mott insulator
phase transition can be captured mathematically by the Bose-Hubbard
Model~\cite{Sachdev}:
\begin{eqnarray}
H_B &=& - J\sum_{\langle i,j\rangle} \left( a^\dagger_i a_j +
a^\dagger_j a_i \right) + U\sum_i n_i(n_i - m), \label{hubbard}
\end{eqnarray}
where $J$ is the tunneling rate between neighbouring lattice sites
$i$ and $j$, and $U$ is the strength of the on-site interactions.
Here, $n_i = a^\dagger_i a_i$ is the number operator at site $i$,
and $m$ is some positive integer number. Initially, the system is in
the superfluid phase which is the ground state of the first $J$-term
of the above Hamiltonian, $|g_I\rangle = \left(\sum_{i=1}^M
a^\dagger_i\right)^K |0\rangle$, where $K$ is the total number of
`atoms' in the superfluid and $|0\rangle$ is the zero-occupation
Fock state.	 This is approximately the coherent state at each site
for large $K$ and large number of sites $M$. A lattice is then
raised adiabatically throughout the superfluid, leading to an
exponential suppression of the tunneling rate $J$ relative to $U$.
At a certain critical value of the ratio $U/J$, the system undergoes
a phase transition to a new state $|g\rangle = \prod_{i=1}^M
\left(a^\dagger_i\right)^m |0\rangle$, which is the ground state of
the second $U$-term in~(\ref{hubbard}).  In considering a single
site $i$ and in the mean field approximation, we replace $a_j$
($j\not = i$) in the Hamiltonian by its mean field $\langle
a_j\rangle \sim \alpha$, which is also a measure of the coherence in
the system. That leads to, up to c-numbers and with some $\alpha$,
\begin{eqnarray}
{\mbox{\rm the $i$-th term in~(\ref{hubbard})}} &\to&
-J\left(a^\dagger_i \langle a_j\rangle + a_i \langle
a^\dagger_j\rangle\right) + U\, n_i(n_i -m)\;,\\
&\to& \tilde J(t)\, (a_i^\dagger - \alpha^*)(a_i - \alpha) + \tilde
U(t) \, (a_i^\dagger a_i - \tilde m)^2 \equiv \tilde J(t)\, H_I +
\tilde U(t)\, H_P\; .\nonumber
\end{eqnarray}
In other words, this is just an implementation of our quantum
adiabatic algorithm for the simple equation $x-\tilde m=0$.
Physically, at each site we have initially the coherent state which
is the ground state of some initial Hamiltonian $H_I$.  At and
beyond the phase transition, the coherence vanishes, $\langle
a_j\rangle = 0$. We are then effectively left with $H_P$, completing
an extrapolation to a final Hamiltonian whose ground state has
$\tilde m$ as the occupation number, which is also the solution for
the simple equation.

\begin{acknowledgments}
I wish to thank Peter Hannaford and Toby Ord for support and
discussions. This work has also been supported by the Swinburne
University Strategic Initiatives.
\end{acknowledgments}

\bibliography{c:/1data_16Apr05/papers/adiabatic}
\bibliographystyle{unsrt}

\end{document}